\documentclass[twocolumn,preprintnumbers,amsmath,amssymb]{revtex4}

\usepackage{graphicx}
\usepackage{dcolumn}
\usepackage{bm}

\begin{document}

\title{Asymmetry of the vapor--liquid coexistence curve:\\
the asymptotic behavior of the ``diameter''}

\author{Miroslav Ya.~Sushko}
\altaffiliation [Corresponding author. \textit{E-mail address}:
]{mrs@onu.edu.ua (M. Ya. Sushko)}
\author{Olexandr M.~Babiy}
\altaffiliation [Present address: ] {Department of Advanced
Mathematics, Odessa National Polytechnic University, 1 Shevchenko
Ave., Odessa 65044, Ukraine.} \affiliation {Department of
Theoretical Physics, Mechnikov
National University, \\
2 Dvoryanska St., Odessa 65026, Ukraine}

\begin{abstract}
\textbf{Abstract}

We analyze, without resort to any model field-mixing scheme, the leading
temperature-dependent term in the `diameter' of the coexistence curve
asymptotically close to the vapor-liquid critical point. For this purpose,
we use a simple non-parametric equation of state which we develop by meeting
several general requirements. Namely, we require that the desired equation
(1) lead to correct asymptotic behavior for a limited number of the fluid's
parameters along selected thermodynamic paths, (2) reveal a Van der Waals
loop below the critical point, and (3) be consistent with a rigorous
definition of the isothermal compressibility in the critical region. For the
temperature interval in question, the proposed equation approximates
experimental data with an accuracy comparable to those given by Schofield's
parametric equation and by other authors' equations. The desired term is
obtained by applying the Maxwell rule to the equation and can be represented
as $D_{2\beta}  {\left| {\tau}  \right|}^{2\beta} $, where ${\left| {\tau}
\right|} = {{{\left| {T - T_{c}}  \right|}} \mathord{\left/ {\vphantom
{{{\left| {T - T_{c}}  \right|}} {T_{c}}} } \right.
\kern-\nulldelimiterspace} {T_{c}}} $ and $\beta $ is the critical exponent
for the order parameter. The amplitude $D_{2\beta}  $ is determined
explicitly for the volume--temperature and entropy--temperature planes.
\bigskip


\end{abstract}

\maketitle
\bigskip
\section{Introduction}

The liquid--vapor critical point has recently attracted considerable
interest in the context of liquid--vapor asymmetry in near-critical fluids
\cite{1,2,3,4}. This asymmetry manifests itself as a deviation of the temperature
behavior of the "diameter" of the coexistence curve, $d \equiv {{(\rho _{1}
+ \rho _{2} )} \mathord{\left/ {\vphantom {{(\rho _{1} + \rho _{2} )} {2\rho
_{c}}} } \right. \kern-\nulldelimiterspace} {2\rho _{c}}} $, from the
critical isochore $\rho = \rho _{c} $. Here $\rho _{1} $ and $\rho _{2} $
are the densities of liquid and saturated vapor for a given temperature $T$,
and $\rho _{c} $ is the critical density. The first empirical evidence of
this fact was obtained by Cailletet and Mathias \cite{5}, who suggested a linear
behavior of $d$ with $T$ as the critical temperature $T_{c} $ is approached:
$d = 1 + D_{1} {\left| {\tau}  \right|}$, where $\tau \equiv {{(T - T_{c} )}
\mathord{\left/ {\vphantom {{(T - T_{c} )} {T_{c}}} } \right.
\kern-\nulldelimiterspace} {T_{c}}} $ is the reduced distance from $T_{c} $
and $D_{1} $ is a constant. This relation is known as the ``law'' of the
rectilinear diameter. The later theoretical investigations \cite{6,7,8} and then
\cite{1,2,3,4} revealed certain deviations from this law, which can be described
with two additional nonanalytic terms, proportional to ${\left| {\tau}
\right|}^{1 - \alpha} $ \cite{6,7,8} and ${\left| {\tau}  \right|}^{2\beta} $
\cite{1,2,3,4}, $\alpha $ and $\beta $ being the critical exponents for the heat
capacity at constant volume $c_{V} $ and the order parameter respectively:
\begin{equation}
\label{eq1}
d = 1 + D_{2\beta}  {\left| {\tau}  \right|}^{2\beta}  + D_{1 - \alpha}
{\left| {\tau}  \right|}^{1 - \alpha}  + D_{1} {\left| {\tau}  \right|} +
... \,.
\end{equation}

For fluids, $1 - \alpha > 2\beta $, and the first temperature-dependent term
on the right of Eq.~(\ref{eq1}) is expected to dominate in the region asymptotically
close to the critical point. Since the extrapolation of $d$ to the critical
temperature is commonly used to obtain the critical density, the
determination of this term becomes of great practical importance.

Theoretically, Eq.~(\ref{eq1}) is usually justified by starting from scaling theory,
dealing with two independent scaling fields, $h_{1} $ and $h_{2} $, and two
conjugate scaling densities, $A_{1} $ and $A_{2} $. The critical part of the
thermodynamic potential is taken to be a homogeneous function of $h_{1} $
and $h_{2} $. The form of the scaling function and the values of $\alpha $
and $\beta $ for a particular model can, in principle, be calculated by using
the renormalization group approach and $\varepsilon $--expansion. For all
models of the same class of universality, these quantities are identical. In
particular, fluids are often treated in terms of the lattice-gas model,
which is isomorphic to the Ising model. However, this statement is
approximate, for the Ising model has perfect symmetry with respect to the
change of sign of the quantities $h_{1} $ and $A_{1} $, which real fluids
lack. As a result, the coexistence curve within this model is symmetrical
(the constants $D$ in Eq.~(\ref{eq1}) are zero). In order to describe the critical
point of real fluids and account for asymmetry of the liquid--vapor
coexistence curve, various field-mixing schemes have been proposed (see Ref.
\cite{9,10} and further developments \cite{1,2}), within which independent physical
fields (say, the chemical potential and temperature) and the conjugate
physical densities (the density and entropy) are represented as linear
combinations of the scaling fields ($h_{1} $ and $h_{2} )$ and densities
($A_{1} $ and $A_{2} )$ of the symmetric model.

Besides solving the problems of the proper choice of the order parameter,
the conjugate field, the form of the scaling function, etc., which are
difficult in themselves, practical implementation of field-mixing schemes
implies the use of certain model assumptions of both the form of the mixing
and the numerical values of the asymmetry coefficients in the corresponding
linear combinations. In practice, these coefficients and, consequently, the
coefficients $D$ in Eq.~(\ref{eq1}) serve as adjustable parameters. The extraction of
their numerical values from experimental data is a very difficult task in
itself (see the discussion of this matter in \cite{4}), due to the difficulties
associated with carrying out experiment in the critical region and the
problem of separation of different terms in Eq.~(\ref{eq1}).

In view of the above-said, the question arises of whether there is an
alternative way of justifying Eq.~(\ref{eq1}), which would allow definite
conclusions on the universality of separate terms in Eq. (\ref{eq1}) and estimations
of their coefficients. In the present communication, we report the results
of study of the leading temperature-dependent terms in $d$ which were
obtained without resort to any model field-mixing scheme. Namely, we proceed
from the idea \cite{11} that an efficient equation of state for the asymptotic
neighborhood of the liquid--vapor critical point can be developed by using
information on the asymptotic behavior of a limited number of the parameters
of the fluid along selected thermodynamic paths. The original approach \cite{11}
is modified in several directions. Firstly, we expand the set of the fluid's
parameters used to construct the desired equation. Secondly, we take into
account a rigorous definition of the isothermal compressibility in the
critical region; this definition becomes a source of asymmetry of the
coexistence curve. Thirdly, we require that for $\tau < 0$, the desired
equation reveal a typical Van der Waals loop. Having met these requirements,
we obtain the desired equation and then test it for the accuracy of
interpolation of experimental data. The sought-for asymptote of $d$ is
determined in the explicit form by applying Maxwell's rule to the equation.
Finally, we analyze the behavior of the diameter in the entropy--temperature
plane.

\section{Construction of the equation of state}

In his work \cite{11}, Martynov developed a simple equation of state for critical
fluid by using the asymptotic law for the behavior of the isothermal
compressibility $\beta _{T} $ on the critical isochore and the asymptotic
form of the equation for the critical isotherm. In the standard variables
$\pi = {{P} \mathord{\left/ {\vphantom {{P} {P_{c}}} } \right.
\kern-\nulldelimiterspace} {P_{c}}}  - 1$, $\tau = {{T} \mathord{\left/
{\vphantom {{T} {T_{c}}} } \right. \kern-\nulldelimiterspace} {T_{c}}}  -
1$, $\omega = {{V} \mathord{\left/ {\vphantom {{V} {V_{c}}} } \right.
\kern-\nulldelimiterspace} {V_{c}}}  - 1$ and for $\tau > 0$, these
relations are written as
\begin{equation}
\label{eq2}
\beta _{T} = {\frac{{1}}{{P_{c}}} }\Gamma _{0} \tau ^{ - \gamma}  + ...\,\,\,  \textrm{as}\,\,\,
\omega = 0 \,\,\, \textrm{and} \,\,\, \tau \to 0^{ +},
\end{equation}
\begin{equation}
\label{eq3}
\pi = - D_{0} \omega \vert \omega \vert ^{\delta - 1} + ...\,\,\,  \textrm{as}\,\,\, \tau = 0.
\end{equation}

The expression for pressure obtained in Ref. \cite{11} fails to adequately
interpolate $P\rho T$ data near the liquid--vapor critical point (see Ref.
\cite{12}). Nonetheless, its implicit functional form was used by the authors of
Ref. \cite{12} as the basis for a scaling equation of state. Invoking the
field-mixing scheme \cite{10}, they developed a nonparametric equation that
interpolates experimental data with an accuracy comparable with that given
by Schofield's parametric equation \cite{13}. However, the expression for $d$
obtained in Ref. \cite{12} contains only one nonanalytic term $ \propto {\left|
{\tau}  \right|}^{1 - \alpha} $.

In the present work, we expand the original set~(2), (3) to incorporate the
well-known fact that the derivative $\left( {{{\partial \pi}
\mathord{\left/ {\vphantom {{\partial \pi}  {\partial \tau}} } \right.
\kern-\nulldelimiterspace} {\partial \tau}} } \right)_{\omega}  $ remains
finite along the critical isochore, including the critical point itself
(see, for instance, Ref. \cite{10}). Consequently, we additionally require that
desired equation satisfy the relation
\begin{equation}
\label{eq4}
\left( {{\frac{{\partial \pi}} {{\partial \tau}} }} \right)_{\omega}  = M =
{\rm c}{\rm o}{\rm n}{\rm s}{\rm t} \,\,\, \textrm{as} \,\,\, \tau = 0 \,\,\, \textrm{and}\,\,\, \omega = 0.
\end{equation}

Also, we pay attention to the fact that the rigorous definition of $\beta
_{T} $ in terms of $\pi $, $\tau $, and $\omega $ is written as
\begin{eqnarray}
\beta _{T} = - {\frac{{1}}{{V}}}\left( {{\frac{{\partial V}}{{\partial P}}}}
\right)_{NT} = - {\frac{{1}}{{1 + \omega}} }{\frac{{1}}{{P_{c}}} }\left(
{{\frac{{\partial \omega}} {{\partial \pi}} }} \right)_{\tau} \nonumber \\ = -
{\frac{{1}}{{P_{c}}} }\left( {{\frac{{\partial \ln (1 + \omega )}}{{\partial
\pi}} }} \right)_{\tau}  .\label{eq5}
\end{eqnarray}
Eq.~(\ref{eq5}) is asymmetric with respect to the transformation $\omega \to -
\omega $, but this source of asymmetry, to our best knowledge, has never
been given an in-depth analysis.

With postulates~(\ref{eq2})--(\ref{eq5}), we begin our consideration with the region $\tau >
0$. Following reasoning \cite{11}, we note that Eq.~(\ref{eq3}) is actually the equation
of state valid only for the curve $\tau = 0$. As we shift from the latter,
the desired equation takes the form $\pi (\tau ,\omega ) = \pi (0,\omega ) +
f(\tau ,\omega )$, where the unknown function $f(\tau ,\omega )$ is expected
to vanish at $\tau = 0$ and to satisfy Eqs.~(\ref{eq2}), (\ref{eq4}) and (\ref{eq5}). It follows
that
\begin{equation}
\label{eq6}
\pi (\tau ,\omega ) = M\tau - {\frac{{1}}{{\Gamma _{0}}} }\ln \left( {1 +
\omega}  \right)\,\tau ^{\gamma}  - D_{0} \omega \vert \omega \vert ^{\delta
- 1},
\quad
\tau > 0.
\end{equation}
For the classical values of the exponents $\gamma = 1$, $\delta = 3$ and
amplitudes $\Gamma _{0} = 1 / 6$, $D_{0} = 3 / 2$, and for the value $M = 4$
Eq.~(\ref{eq6}) coincides with the asymptotic form of the Van der Waals equation for
$\tau ,\,\omega < < 1$: $\pi = 4\tau - 6\omega \tau - (3 / 2)\omega ^{3} +
...\,$.

Next, we suggest that below the critical temperature ($\tau = - \vert \tau
\vert < 0)$, Eq.~(\ref{eq6}) must be generalized so as to demonstrate a typical Van
der Waals loop and transform, for the corresponding values of the exponents
and amplitudes, to the asymptotic form of the Van der Waals equation for
$\vert \tau \vert ,\,\omega < < 1$: $\pi = - 4{\left| {\tau}  \right|} +
6\omega {\left| {\tau}  \right|} - (3 / 2)\omega ^{3} + ...\,$. Then
\begin{equation}
\label{eq7}
\pi (\tau ,\omega ) = - M{\left| {\tau}  \right|} + {\frac{{1}}{{\Gamma _{0}
}}}\ln (1 + \omega )\,{\left| {\tau}  \right|}^{\gamma}  - D_{0} \omega
{\left| {\omega}  \right|}^{\delta - 1},
\quad
\tau < 0.
\end{equation}

Combined together, Eqs.~(\ref{eq6}) and (\ref{eq8}) give
\begin{equation}
\label{eq8}
\pi (\tau ,\omega ) = M\tau - {\frac{{1}}{{\Gamma _{0}}} }\ln (1 + \omega
)\tau \,{\left| {\tau}  \right|}^{\gamma - 1} - D_{0} \omega {\left| {\omega
} \right|}^{\delta - 1}.
\end{equation}
Eq.~(\ref{eq8}) represents the desired equation and forms the basis for further
analysis.

\section{Testing the equation of state}

In the quadratic approximation in $\omega $, $\ln (1 + \omega )
\approx \omega - \omega ^{2} / 2$, and Eq.~(\ref{eq8}) almost
coincides with the equation obtained in Ref.~\cite{12}, except for
terms of order $\vert \omega \vert ^{\delta + 1}$ and $\vert \tau
\vert ^{2 - \alpha} $. In its accuracy, the latter proves to be
comparable with Schofield's parametric equation.

Here, we consider a more sophisticated way of testing Eq.~(\ref{eq8}), hinted by the
suggestion that the law of corresponding states can be extended to the
asymptotic neighborhood of the liquid--vapor critical point. If so, then the
variables $\pi $, $\tau $, and $\omega $ can be properly normalized so as to
fall on one and the same curve. For Eq.~(\ref{eq8}), the form of this curve is
particularly suitable for processing, with only one adjustable parameter
actually used.

The presence of the logarithmic factor in Eq.~(\ref{eq8}) makes the choice of the
normalized variables simple. After we set
\begin{equation}
\label{eq9}
p = {\frac{{\pi}} {{D_{0}}} },
\quad
t = {\frac{{\tau}} {{(D_{0} \Gamma _{0} )^{1 / \gamma}} }},
\quad
v = \omega ,
\end{equation}
Eq.~(\ref{eq8}) takes the form
\begin{equation}
\label{eq10} p(t,v) = at - \ln \left( {1 + v} \right)
t\,|t|^{\gamma-1} - v |v|^{\delta-1} ,
\end{equation}
\noindent
where $a$ is a constant. If the law of corresponding states does occur for a
certain class of near-critical fluids, $a$ remains the same for all members
of the class. In particular, the Van der Waals and Dieterici equations give
estimates $a = 2 / 3$ and $a = 3 / 4$ respectively.

To test the validity of Eq.~(\ref{eq10}) for interpolation of $P\rho T$ data in the
close vicinity of the critical point, we used data for water \cite{14,15},
nitrogen \cite{16}, and carbon dioxide \cite{17}. The values of the critical exponents
for all the substances were taken to be equal to their values in the
three-dimensional Ising model: $\gamma = 1.239$, $\delta = 4.80$ \cite{18}. The
critical amplitudes were calculated by using the values of the parameter $z
= {{P_{c} V_{c}}  \mathord{\left/ {\vphantom {{P_{c} V_{c}}  {RT_{c}}} }
\right. \kern-\nulldelimiterspace} {RT_{c}}} $ and approximation formulas
\cite{19}. The parameters of the critical points were taken from the original
works \cite{14,15,16,17}.

To scrutinize the functional form of Eq.~(\ref{eq10}), the values
of the function $f = p + \ln \left( {1 + v} \right)t
\,|t|^{\gamma-1} + v |v|^{\delta-1} $ were plotted for available
values of the variables (\ref{eq9}), and then the possibility of
their interpolation with a straight line $at$ was analyzed (see
Figs. 1 and 2, where $\Delta \tilde {\rho}  \equiv {{\left( {\rho
- \rho _{c}} \right)} \mathord{\left/ {\vphantom {{\left( {\rho -
\rho _{c}} \right)} {\rho _{c} }}} \right.
\kern-\nulldelimiterspace} {\rho _{c}}} )$. The $f$ versus $t$
dependence was indeed found to approach a linear one as the
density interval was narrowed. The slope $a$ of the interpolation
line was determined with the least square procedure and then used
to estimate the root-mean-square error $\Delta_P = \sqrt {N^{ -
1}{\sum\limits_{1}^{N} {{\left[ {\left( {P_{{\rm e}{\rm x}{\rm p}}
- P_{{\rm t}{\rm h}{\rm e}{\rm o}{\rm r}}} \right) / P_{{\rm
e}{\rm x}{\rm p}}}  \right]^{2}}}} } $ for direct approximation of
the $P\rho T$-data with the original equation (\ref{eq8}) (see
Fig. 3). In all cases studied, $\Delta_P$ was comparable with that
obtained in Ref. \cite{12}, and even lower for water.

\bigskip
\begin{figure}
\centering
\includegraphics[width=8.6cm]{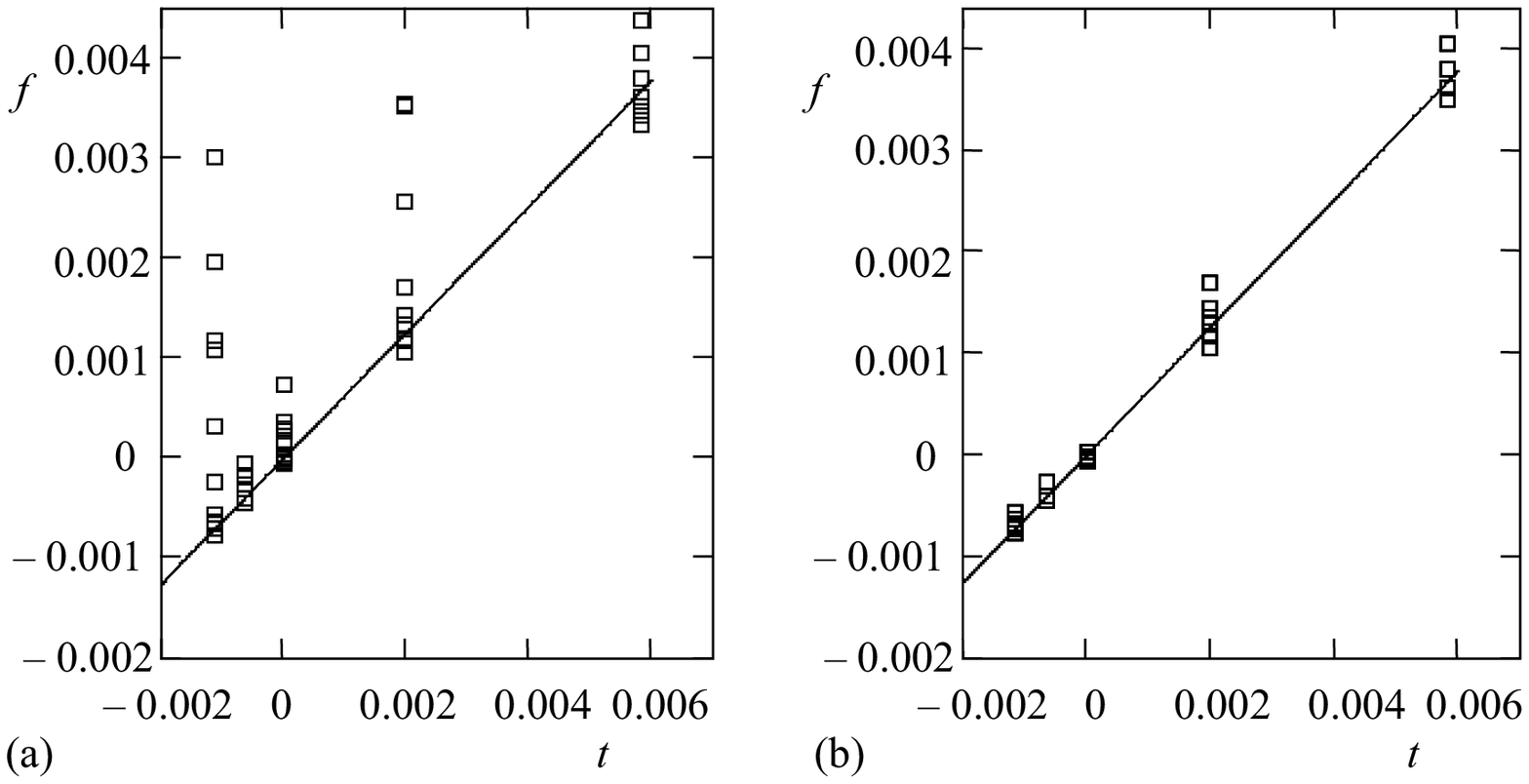}
\caption{$f$ versus $t$ in the region $ - {\rm 2}.{\rm 4}{\rm
7}{\rm 2}\times {\rm 1}{\rm 0}^{{\rm -} {\rm 4}} < \tau < {\rm
1}.{\rm 2}{\rm 3}{\rm 6}\times {\rm 1}{\rm 0}^{{\rm -} {\rm 3}}$,
$ - 0.3 \le \Delta \tilde {\rho}  \le 0.3$ according to data
\cite{14,15} for water (a) and its asymptotic behavior for $ -
0.15 \le \Delta \tilde {\rho}  \le 0.15$ (b); $a = 0.63$ and
$\Delta_P = 0.27\% $. }
\end{figure}
\begin{figure}
\centering
\includegraphics[width=8.6cm]{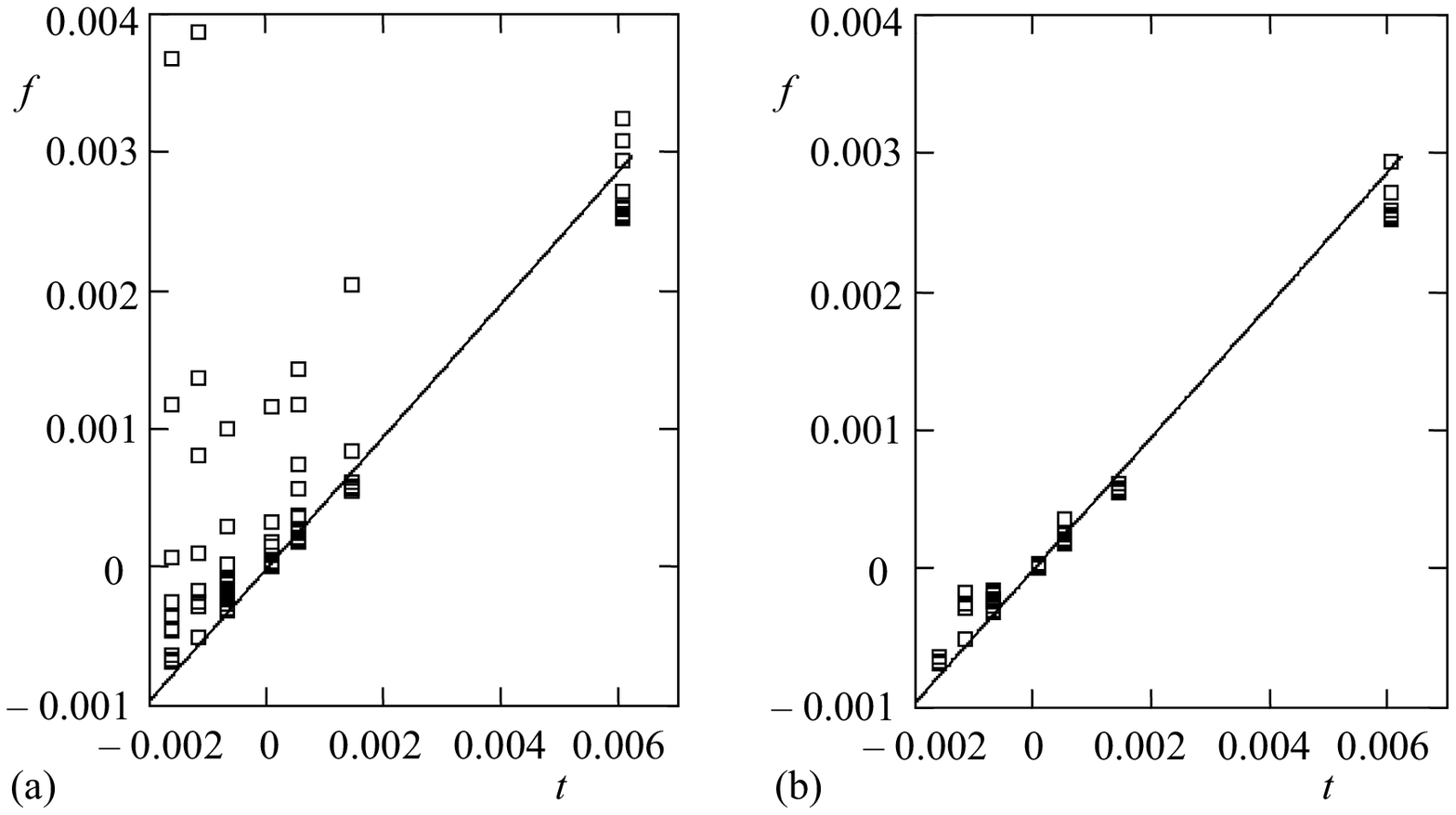}
\caption{$f$ versus $t$ in the region $ - {\rm 1}.{\rm 2}{\rm
3}{\rm 6}\times {\rm 1}{\rm 0}^{{\rm -} {\rm 3}} < \tau < {\rm
3}\times {\rm 1}{\rm 0}^{{\rm -} {\rm 3}}$, $ - 0.23 \le \Delta
\tilde {\rho}  \le 0.23$ according to data \cite{16} for nitrogen
(a) and its asymptotic behavior for $ - 0.15 \le \Delta \tilde
{\rho}  \le 0.15$ (b); $a = 0.48$ and $\Delta_P = 0.4\% $.}
\end{figure}
\begin{figure}
\centering
\includegraphics[width=7cm]{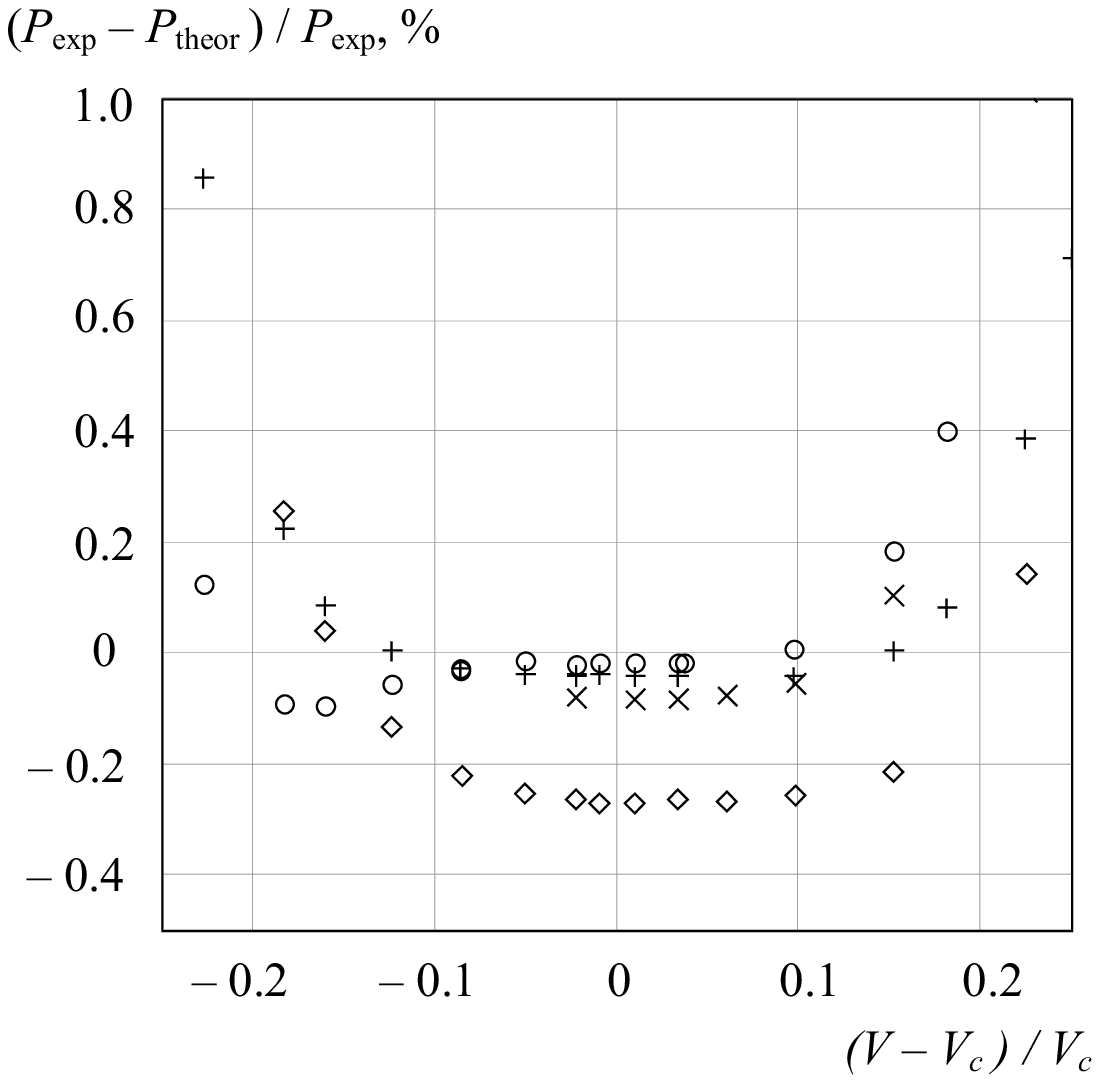}
\caption{Deviation $\left( {P_{\exp}  - P_{{\rm t}{\rm h}{\rm e}{\rm o}{\rm
r}}}  \right) / P_{\exp}  \cdot 100\% $ of data \cite{16} from Eq.~(\ref{eq8}), with $a =
0.48$, for isotherms $\tau = 3.169\times 10^{ - 5}$ ($\circ$), $2.{\rm
6}{\rm 9}{\rm 4}\times {\rm 1}{\rm 0}^{{\rm -} {\rm 4}}$ (+), $7.{\rm
4}{\rm 4}{\rm 8}\times {\rm 1}{\rm 0}^{{\rm -} {\rm 4}}$ ($\times$), and
${\rm 3}.{\rm 1}{\rm 2}{\rm 2}\times {\rm 1}{\rm 0}^{{\rm -} {\rm 3}}$
($\diamond$).}
\end{figure}

\section{The coexistence curve}

Let $V_{1} $ and $V_{2} $ be the maximum and minimum molar volumes (for a
given $\tau )$ of the coexisting liquid and vapor respectively, and let
$\omega _{1} = {{\left( {V_{1} - V_{c}}  \right)} \mathord{\left/ {\vphantom
{{\left( {V_{1} - V_{c}}  \right)} {V{}_{c}}}} \right.
\kern-\nulldelimiterspace} {V{}_{c}}} = - {\left| {\omega _{1}}  \right|}$
and $\omega _{2} = {{\left( {V_{2} - V_{c}}  \right)} \mathord{\left/
{\vphantom {{\left( {V_{2} - V_{c}}  \right)} {V{}_{c}}}} \right.
\kern-\nulldelimiterspace} {V{}_{c}}} = \omega _{2} > 0$ be the
corresponding values of $\omega $. It is well-known (see, for example, \cite{20})
that the form of the coexistence curve can be determined from the equality
of the chemical potentials for the coexisting phases, or, equivalently, from
Maxwell's rule.

Below the critical point, the liquid ($\omega \le \omega _{1} )$ and vapor
($\omega \ge \omega _{2} )$ phases satisfy Eq.~(\ref{eq8}). In particular, for the
values $\omega _{1} $ and $\omega _{2} $ on the coexistence curve we have:
\begin{equation}
\label{eq11} \pi (\tau ,\omega _{1} ) = - M{\left| {\tau} \right|}
+ {\frac{{1}}{{\Gamma _{0}}} }\ln \left( {1 - {\left| {\omega
_{1}}  \right|}} \right)\,{\left| {\tau} \right|}^{\gamma}  +
D_{0} \,{\left| {\omega _{1}} \right|}^{\delta},
\end{equation}
\begin{equation}
\label{eq12} \pi (\tau ,\omega _{2} ) = - M{\left| {\tau} \right|}
+ {\frac{{1}}{{\Gamma _{0}}} }\ln \left( {1 + \omega _{2}}
\right)\,{\left| {\tau} \right|}^{\gamma}  - D_{0} \,{\left|
{\omega _{2}} \right|}^{\delta}.
\end{equation}

For the two-phase region $(\omega _{1} ,\omega _{2} )$, Eq.~(\ref{eq8}) must be
corrected according to Maxwell's rule:
\begin{equation}
\label{eq13}
{\int\limits_{\omega _{1}} ^{\omega _{2}}  {\pi (\tau ,\omega )\,d\omega =
\bar {\pi} \left( {\omega _{2} - \omega _{1}}  \right)}} ,
\end{equation}
\noindent
where $\bar {\pi} $ is the constant pressure along the isotherm--isobar
$(\omega _{1} ,\omega _{2} )$. At the endpoints of this segment, $\bar {\pi
}$ is equal to the values given by (\ref{eq11}), (\ref{eq12}): $\bar {\pi}  = \pi (\tau
,\omega _{1} ) = \pi (\tau ,\omega _{1} )$.

Taking the integral (\ref{eq13}) and combining the result with Eqs.~(\ref{eq11}) and (\ref{eq12}),
we find:
\begin{equation}
\label{eq14}
{\frac{{1}}{{\Gamma _{0}}} }{\left| {\tau}  \right|}^{\gamma} {\left[ {\ln
\left( {1 + \omega _{2}}  \right) - \ln \left( {1 - {\left| {\omega _{1}}
\right|}} \right)} \right]} = D_{0} \left( {\omega _{2} ^{\delta}  + {\left|
{\omega _{1}}  \right|}^{\delta}}  \right)\,,
\end{equation}
\begin{eqnarray}
{\frac{{1}}{{\Gamma _{0}}} }\,{\left| {\tau}  \right|}^{\gamma} \left(
{\omega _{2} + {\left| {\omega _{1}}  \right|}} \right) = {\frac{{\delta
}}{{\delta + 1}}}D_{0} \,\left( {\omega _{2} ^{\delta + 1} - {\left| {\omega
_{1}}  \right|}^{\delta + 1}} \right) \nonumber\\ + D_{0} \left( {\omega _{2} ^{\delta}
+ {\left| {\omega _{1}}  \right|}^{\delta}}  \right). \label{eq15}
\end{eqnarray}

Thus, the analysis of the coexistence curve reduces to the study of the
system of Eqs.~(\ref{eq14}), (\ref{eq15}). To obtain the shape of this curve, we define a
nonnegative quantity $x$ by the relation
\begin{equation}
\label{eq16}
\omega _{2} = x{\left| {\omega _{1}}  \right|},
\end{equation}
\noindent
and then divide both sides of Eqs.~(\ref{eq14}), (\ref{eq15}) into each other to eliminate
the variable $\tau $. As a result, we see that $x$ is the only root of the
transcendental equation
\begin{eqnarray}
{\left[ {{\frac{{\delta}} {{\delta + 1}}}{\left| {\omega _{1}}
\right|}\,\left( {x^{\delta + 1} - 1} \right) + \left( {x^{\delta}  + 1}
\right)} \right]}\,\ln {\frac{{1 + {\left| {\omega _{1}}  \right|}x}}{{1 -
{\left| {\omega _{1}}  \right|}}}}\nonumber \\ - {\left| {\omega _{1}}  \right|}\left(
{x^{\delta}  + 1} \right)\left( {x + 1} \right) = 0, \label{eq17}
\end{eqnarray}
\noindent with ${\left| {\omega _{1}}  \right|}$ being a
parameter. For a fixed value of ${\left| {\omega _{1}}  \right|}$
and near the point $x=1$, the left side of Eq.~(\ref{eq17}) is a
monotone increasing function of $x$. The single root of this
function is close to unity and is located in the region $x < 1$.

To obtain the analytical solution of the system (\ref{eq14}), (\ref{eq15}), we represent $x
= 1 + \varepsilon $, where $\varepsilon < < 1$. In the linear approximation
with respect to $\varepsilon $, Eq.~(\ref{eq17}) gives
\begin{equation}
\label{eq18} \varepsilon =  {\frac{{4{\left| {\omega _{1}}
\right|}\left( {1 - {\frac{{1}}{{2{\left| {\omega _{1}}
\right|}}}}\ln {\frac{{1 + {\left| {\omega _{1}}  \right|}}}{{1 -
{\left| {\omega _{1}}  \right|}}}}} \right)}}{{\delta \left( {1 +
{\left| {\omega _{1}}  \right|}} \right)\ln {\frac{{1 + {\left|
{\omega _{1}}  \right|}}}{{1 - {\left| {\omega _{1}} \right|}}}} -
2{\left| {\omega _{1}}  \right|}\left( {\delta + 1 -
{\frac{{1}}{{1 + {\left| {\omega _{1}}  \right|}}}}} \right)\,}}}.
\end{equation}

Expanding the logarithmic functions on the right of Eq.~(\ref{eq18}) in power series
with respect to ${\left| {\omega _{1}}  \right|}$ and restricting ourselves
to the linear approximation, we find $\varepsilon = 0$ and $\omega _{2} =
{\left| {\omega _{1}}  \right|}$. Eq.~(\ref{eq14}) then immediately gives the
well-known asymptotic law
\begin{equation}
\label{eq19}
\omega _{2} = - \omega _{1} = B_{0} {\left| {\tau}  \right|}^{\beta}  +
o\left( {{\left| {\tau}  \right|}^{\beta}}  \right),
\end{equation}
\noindent
with amplitude
\begin{equation}
\label{eq20}
B_{0} = {\frac{{1}}{{\left( {D_{0} \Gamma _{0}}  \right)^{{{1}
\mathord{\left/ {\vphantom {{1} {(\delta - 1)}}} \right.
\kern-\nulldelimiterspace} {(\delta - 1)}}}}}}.
\end{equation}

In the second-order approximation with respect to ${\left| {\omega _{1}}
\right|}$, these two relations still hold. Asymmetry of the coexistence
curve appears only within the third-order approximation, in which case we
find that asymptotically close to the critical point,
\begin{equation}
\label{eq21}
\varepsilon = - {\frac{{2{\left| {\omega _{1}}  \right|}}}{{3\left( {\delta
- 1} \right)}}} + O\left( {{\left| {\omega _{1}}  \right|}^{2}} \right){\rm
,}
\end{equation}
\begin{equation}
\label{eq22}
\omega _{2} - \omega _{1} = 2B_{0} {\left| {\tau}  \right|}^{\beta}  +
O\left( {{\left| {\tau}  \right|}^{2\beta}}  \right),
\end{equation}
\begin{equation}
\label{eq23}
\omega _{2} + \omega _{1} = \varepsilon {\left| {\omega _{1}}  \right|} = -
{\frac{{2}}{{3\left( {\delta - 1} \right)}}}B_{0}^{2} {\left| {\tau}
\right|}^{2\beta}  + o\left( {{\left| {\tau}  \right|}^{2\beta}}
\right).
\end{equation}
It follows from Eq.~(\ref{eq23}) that the desired asymptotic expression for the
``diameter'' of the coexistence curve in the volume--temperature plane is
\begin{equation}
\label{eq24}
d_{V} = {\frac{{V_{1} + V_{2}}} {{2V_{c}}} } = 1 - {\frac{{1}}{{3\left(
{\delta - 1} \right)}}}\,B_{0}^{2} {\left| {\tau}  \right|}^{2\beta}  +
o\left( {{\left| {\tau}  \right|}^{2\beta}}  \right).
\end{equation}

\section{The entropy--temperature plane}

Incorporating an additional requirement \cite{11} on the asymptotic behavior of
the molar heat capacity $c_{V} (\tau ,\omega )$ along the critical isochore
$\omega = 0$ in the two-phase region,
\begin{equation}
\label{eq25} c_{V} (\tau ,0) = A_{0} {\left| {\tau}  \right|}^{ -
\alpha}  + ...\,\,\, \textrm{as} \,\,\, \tau \to 0^{ -},
\end{equation}
\noindent
let us determine the ``diameter'' $d_{s} = {{\left( {s_{1} + s_{2}}
\right)} \mathord{\left/ {\vphantom {{\left( {s_{1} + s_{2}}  \right)}
{s_{c}}} } \right. \kern-\nulldelimiterspace} {s_{c}}} $ of the coexistence
curve in the entropy--temperature plane. For this purpose, we first use the
relation $d\mu = - sdT + VdP$, $s$ and $V$ being the molar entropy and
volume of the system, to find the chemical potential $\mu $. In terms of
$\tau $, $\omega $, and $\pi $, and for a constant temperature, we have
\begin{equation}
\label{eq26}
(d\mu )_{\tau}  = V_{c} P_{c} (1 + \omega )\,(d\pi )_{\tau}  .
\end{equation}
Correspondingly,
\begin{equation}
\label{eq27}
\mu (\tau ,\omega ) = V_{c} P_{c} {\left( {\pi + \int {\omega}  \,d\pi}
\right)} + f(\tau ),
\end{equation}
\noindent
where the integral is taken along an isotherm and $f(\tau )$ is a function
of temperature alone. In view of Eq.~(\ref{eq8}), we find
\begin{eqnarray}
\mu (\tau ,\omega ) = P_{c} V_{c} {\left[ {M\tau - {\frac{{1}}{{\Gamma _{0}
}}}\omega \,\tau \,{\left| {\tau}  \right|}^{\gamma - 1}} \right.}\nonumber \\
{\left. - D_{0}  {\omega} \,{\left| {\omega} \right|}^{\delta - 1}
- D_{0} {\frac{{\delta}} {{\delta + 1}}} \, {\left| {\omega}
\right|}^{\delta + 1} \right]} + f(\tau ). \label{eq28}
\end{eqnarray}

In the asymptotic vicinity of the critical point, the fourth term in the
brackets can be neglected, and Eq.~(\ref{eq28}) takes the form
\begin{equation}
\label{eq29} {\frac{{\mu (\tau ,\omega ) - \mu (\tau ,0)}}{{P_{c}
V_{c}}} } = - \omega {\left| {\omega}  \right|}^{\delta -
1}h\left( {{{\tau}  \mathord{\left/ {\vphantom {{\tau}  {{\left|
{\omega}  \right|}^{\beta}} }} \right. \kern-\nulldelimiterspace}
{{\left| {\omega}  \right|}^{1/\beta}} }} \right),
\end{equation}
\noindent
in accordance with the scaling hypothesis \cite{21}. The asymptote of the scaling
function $h(x)$ for $x < < 1$ is $h(x) = D_{0} + {\frac{{1}}{{\Gamma _{0}
}}}x^{\gamma} $. Implicitly present in Ref. \cite{11}, this form of the scaling
function was later postulated in Ref. \cite{12}. The omitted term in the brackets
in Eq.~(\ref{eq28}) represents, evidently, the first correction to the asymptotic
law (\ref{eq29}).

As a check on the validity of Eq.~(\ref{eq28}), we can easy verify that the
phase-equilibrium conditions for two coexisting phases with molar volumes
$\omega _{1} = - {\left| {\omega _{1}}  \right|}$ and $\omega _{2} $ ($\tau
< 0)$,
\begin{equation}
\label{eq30}
\pi (\tau ,\omega _{1} ) = \pi (\tau ,\omega _{2} ),
\quad
\mu (\tau ,\omega _{1} ) = \mu (\tau ,\omega _{2} ),
\end{equation}
\noindent
reduce to Eq.~(\ref{eq14}) and (\ref{eq15}), provided, as was already suggested, that
$f(\tau )$ is independent of $\omega $.

In view of Eq.~(\ref{eq28}), the molar entropy,
\begin{equation}
\label{eq31}
s = - {\frac{{1}}{{T_{c}}} }\left( {{\frac{{\partial \mu}} {{\partial \tau
}}}} \right)_{\pi}  = {\frac{{P_{c} V_{c}}} {{T_{c}}} }(1 + \omega )\left(
{{\frac{{\partial \pi}} {{\partial \tau}} }} \right)_{\omega}  -
{\frac{{1}}{{T_{c}}} }\left( {{\frac{{\partial \mu}} {{\partial \tau}} }}
\right)_{\omega},
\end{equation}
\noindent
is
\begin{eqnarray}
s(\tau ,\omega) = {\frac{{P_{c} V_{c}}} {{T_{c}}} } {\left\{ M\omega +
{\frac{{1}}{{\Gamma _{0}}} } \left[ {\omega - (1 + \omega )\ln \left( {1 +
\omega}  \right)} \right] \right.} \nonumber \\
{\left. \times \gamma\,  \,{\left| {\tau} \right|}^{\gamma - 1}
\right\}}- {\frac{{1}}{{T_{c}}} }{f}'(\tau). \label{eq32}
\end{eqnarray}
Correspondingly, the molar heat capacity at constant volume,
\begin{equation}
\label{eq33}
c_{V} = T\left( {{\frac{{\partial s}}{{\partial T}}}} \right)_{V} = (1 +
\tau )\left( {{\frac{{\partial s}}{{\partial \tau}} }} \right)_{\omega}
,
\end{equation}
\noindent
is given by
\begin{eqnarray}
c_{V} (\tau ,\omega ) = {\frac{{P_{c} V_{c}}} {{T_{c}}} }{\frac{{1}}{{\Gamma
_{0}}} }{\left[ \,{\omega - (1 + \omega )\ln \left( {1 + \omega}  \right)}
\right]}\,\nonumber \\
\times \gamma (\gamma - 1)\textrm{sgn}(\tau )(1 + \tau )\,{\left| {\tau}
\right|}^{\gamma - 2} - {\frac{{1 + \tau}} {{T_{c}}} }{f}''(\tau ).
\label{eq34}
\end{eqnarray}

If, according to Eq.~(\ref{eq25}),
\begin{equation}
\label{eq35}
 \frac{P_c V_c}{T_c} \frac{\gamma(\gamma-1)B_0^2}{2\Gamma_0} (1+\tau)
  |\tau|^{-\alpha}- {\frac{{1 + \tau}} {{T_{c}}} }{f}''(\tau ) = A_{0} {\left| {\tau}
\right|}^{ - \alpha}  + ...\, ,
\end{equation}
\noindent
then, taking into account that the critical value of the entropy $s_{c} =
s(0,0) = - {{{f}'(0)} \mathord{\left/ {\vphantom {{{f}'(0)} {T_{c}}} }
\right. \kern-\nulldelimiterspace} {T_{c}}} $, we find
\begin{eqnarray}
 s(\tau ,\omega ) = {\frac{{P_{c} V_{c}}} {{T_{c}}}}{\left\{ {M\omega +
{\frac{{\gamma}}{{\Gamma _{0}}} }{\left[ \,{\omega - (1 + \omega
)\ln
\left( {1 + \omega}  \right)} \right]}}{\left| {\tau} \right|}^{\gamma - 1} \right\}} \nonumber \\
+ {\int\limits_{0}^{\tau} {d\tau} \left\{\left[\frac{A_0}{1+
\tau}-\frac{P_c V_c}{T_c} \frac{\gamma(\gamma-1)B_0^2}{2\Gamma_0}
\right]\vert \tau \vert ^{ - \alpha} + o\left(\vert \tau \vert ^{
- \alpha}\right)
\right\}}\nonumber\\
+  \,\,s_{c}, \,\,\,\,\,\,\,\,\,\,\,\,\,\,\,\label{eq36}
\end{eqnarray}
\noindent whence
\begin{eqnarray}
d_{s} = 1 - {\frac{{P_{c} V_{c}}} {{T_{c} s_{c}}}
}{\frac{{MB_{0}^{2} }}{{3(\delta - 1)}}}{\left| {\tau}
\right|}^{2\beta} \nonumber \\ - {\frac{{1}}{{s_{c}(1-\alpha)
}}}\left( {A_{0} + {\frac{{P_{c} V_{c}}} {{T_{c} }}}{\frac{{\beta
\gamma B_{0}^{2}}} {{\Gamma _{0}}} }} \right){\left| {\tau}
\right|}^{1 - \alpha}  + o\left( {{\left| {\tau} \right|}^{1 -
\alpha}} \right). \label{eq37}
\end{eqnarray}

It follows that condition (\ref{eq25}) is not necessary to explain
the appearance of the ${\left| {\tau}  \right|}^{2\beta} $ term in
the ``diameter'' (\ref{eq37}). On the other hand, Eq.~(\ref{eq34})
shows that on both branches of the coexistence curve, $c_{V} $
diverges as ${\left| {\tau}  \right|}^{ - \alpha} $ as the
critical point is approached from below.

\section{Conclusion}

The temperature behavior of the ``diameter'' of the coexistence curve in the
asymptotic vicinity of the vapor--liquid critical point has been studied
within a model-free approach, based upon general thermodynamic definitions
and relationships. The critical exponent of the leading
temperature-dependent term in the diameter is found to be $2\beta $. The
critical amplitude for this term is determined explicitly for the
volume--temperature and entropy--temperature planes. In the latter case, the
``$1 - \alpha $'' term has been recovered as well.

\bigskip

\_\_\_\_\_\_\_\_\_\_\_\_\_\_\_\_\_\_\_\_\_\_\_\_\_\_\_\_\_\_\_\_\_\_\_\_\_

This report was presented at a Section III (Phase transitions and critical phenomena) Oral Session of
 4$^{{\rm t}{\rm h}}$ International Conference \textit{Physics of Liquid Matter: Modern Problems},
  23--26 May, 2008, Kyiv, Ukraine. Its full version is published
  in \emph{Journal of Molecular Liquids}.

\end{document}